\documentclass[12pt]{article}
\usepackage{amsmath}
\usepackage{graphicx}
\usepackage{enumerate}
\usepackage{natbib}
\usepackage{comment}
\usepackage{url} 
\usepackage{bm}
\usepackage{amsfonts}
\usepackage[flushleft]{threeparttable}
\usepackage{booktabs}
\usepackage{xcolor}
\usepackage[medium,compact]{titlesec}
\usepackage{tikz}
\usepackage{multirow}
\usepackage{subcaption}
\usepackage{bigints}
\usepackage{amsthm}
\usepackage{hyperref}
\usepackage{paralist}
\usepackage[nodisplayskipstretch]{setspace}
\usepackage{amssymb}
\usepackage{tabularx}
\usepackage{array}

\usepackage{tikz}
\usetikzlibrary{decorations.pathreplacing,decorations.pathmorphing,arrows.meta,positioning,calc}

\allowdisplaybreaks

\newcommand{\blind}{1}

\newtheorem{remark}{Remark}

\newtheorem{assumption}{Assumption}
\newtheorem{theorem}{Theorem}
\newtheorem{proposition}{Proposition}
\newtheorem{corollary}{Corollary}

\theoremstyle{plain}

\newcommand{\indep}{\perp\!\!\!\perp}

\newcommand{\bcx}{{\bm X}}

\newcommand{\bco}{{\bm O}}

\newcommand{\E}{\mathbb{E}}
\newcommand{\Prob}{\mathbb{P}}

\allowdisplaybreaks

\begin{document}

\def\spacingset#1{\renewcommand{\baselinestretch}%
{#1}\small\normalsize} \spacingset{1}


\if1\blind
{
  \title{\bf \Large Empirical stratification for predictive  treatment effect heterogeneity with post-treatment variables}
  \author{Chao Cheng$^{1,*}$ \quad \quad Rui Wang$^{2,\dagger}$\vspace{0.2cm} \quad \quad Yichi Zhang$^{3,\mathsection}$\\
  $^1$Department of Statistics and Data Science, \\
  Washington University in St. Louis, MO\\
  $^2$Department of Resource Economics, \\
  University of Massachusetts Amherst, MA\\
  $^3$Department of Statistics,\\ Indiana University Bloomington, IN \\
    $^*$chaoc@wustl.edu \quad $^\dagger$rwang0@umass.edu \quad  $^\mathsection$yiczhan@iu.edu}
  \maketitle

  \vspace{-1cm} 
  
} \fi

\if0\blind
{
  \bigskip
  \bigskip
  \bigskip
  \begin{center}
    {\Large \bf Empirical stratification for predictive treatment effect heterogeneity with post-treatment variables}
\end{center}
  \medskip
} \fi

\begin{abstract}
\noindent Post-treatment variables (PVs), such as intercurrent events, treatment noncompliance, and behavioral responses to treatment, provide information about individuals' post-treatment responses and may help characterize heterogeneity in treatment effects on the primary outcome.  This paper develops an empirical stratification framework to study  the treatment effect heterogeneity across baseline-predicted PV response profiles. Specifically, we construct empirical scores from baseline-covariate predictions of potential PV responses and use these scores to define empirically accessible subgroups for treatment effect evaluation. The resulting empirical-stratum treatment effects (ETEs) quantify treatment effects across these baseline-predicted PV response profiles, and are identifiable under a standard set of assumptions in causal inference.  We further introduce projected ETE curves and develop efficient-influence-function-based estimators that allow flexible nuisance estimation. We clarify the distinction and connection of our framework to principal stratification analysis, which address two inferential questions. We evaluate the proposed framework through simulation studies and illustrate its use in two real-world applications.
\end{abstract}

\noindent%

\spacingset{1.65} 


\section{Introduction}

\subsection{Motivation and research questions}

Heterogeneous treatment effects (HTEs) are central to many scientific and policy questions. 
Existing HTE methods focus on treatment effect heterogeneity for baseline covariates 
\citep[e.g.,][]{wager2018estimation,semenova2021debiased,kennedy2023towards,chernozhukov2025generic}. 
In many applications, the scientifically relevant subgroup is 
characterized by a post-treatment variable (PV). Examples include intercurrent events by the ICH E9 estimands framework \citep{kahan2024estimands}, such as disease worsening or adverse events in medical settings; treatment noncompliance \citep{angrist1996identification}; and behavioral responses to treatment \citep{manski2013public}. These PVs are scientifically informative on how individuals respond to treatment initiation and therefore may also characterize  treatment effect variation in the primary outcome. 
However, because PVs are themselves affected by treatment, directly conditioning on the observed PV can induce endogenous selection bias 
\citep{elwert2014endogenous,montgomery2018conditioning}. Thus, conventional HTE methods developed for baseline covariates cannot be directly applied to PVs. This motivates considerable interest in causal frameworks to characterize treatment effect heterogeneity related to PVs.

A representative framework for studying HTEs related to PVs is principal stratification \citep{frangakis2002principal}. 
It defines principal causal effects within latent principal strata, which are  subgroups characterized by the joint counterfactual values of the PV under alternative treatment conditions. 
For example, when the PV is an adverse event due to a medical treatment, the principal stratification framework classifies individuals according to whether this  event would occur under each medical treatment condition. It therefore provides a scientifically transparent way to describe treatment effect variations across counterfactual PV response types. 
However, because principal strata are latent, identifying principal causal effects requires strong structural assumptions, such as monotonicity, exclusion restrictions, principal ignorability, and parametric modeling assumptions 
\citep{angrist1996identification,ding2017principal,liu2024principal}. 
These assumptions can be persuadable in certain applications, but they are typically problem-specific and hard to verify, thereby significantly complicating robust inference in principal stratification analysis \citep{vansteelandt2025chasing}.

Alternatively, there exists a more empirical strategy to study the \textit{predictive treatment effect heterogeneity} with subgroups defined by baseline-predicted PV responses. Specifically, baseline covariates are first used to predict the PV responses under each treatment condition, and the resulting prediction scores are then used to stratify individuals for treatment effect analysis.  
For example,   \citet{vanderweele2012conditioning} investigate a perinatal epidemiologic study for the effect of maternal smoking on infant mortality, where the occurrence of low birth weight of infant is the PV of interest. The authors stratify infants by their baseline-predicted probability of low birth weight in order to study the variation of treatment effect across infants with different risks of low birth weight. This idea is commonly referred to as predictive approaches to HTE analysis and has been widely used in the medical literature \citep{kent2018personalized,kent2020predictive}, with a recent review identifying 83 publications conducting such analyses in randomized trials from 2021 through 2024 \citep{selby2025predictive}. 
In the social science literature, this idea is often referred to as endogenous stratification \citep{abadie2018endogenous} and has been applied in several empirical studies; see, for example, \citet{pane2014effectiveness,carranza2020job}.

Despite the considerable attention that predictive treatment effect heterogeneity has received in practical applications, existing approaches have largely been developed outside of the causal inference perspective, such as regression-based analyses in observational studies \citep[e.g.,][]{vanderweele2012conditioning} or treatment effectiveness comparisons in randomized trials \citep[e.g.,][]{kent2018personalized,abadie2018endogenous}. This motivates us to formalize predictive treatment effect heterogeneity  under a causal inference framework. Specifically, we first define subgroups, referred to as empirical strata, using baseline predictions of the potential PV responses under alternative treatment conditions. We then introduce empirical-stratum treatment effects (ETEs) as a new class of HTEs to quantify treatment effects within these empirical strata. For instance, if the PV is an adverse event, the ETEs can help quantify how the treatment effect on the final outcome varies across these different predicted-risk of experiencing such adverse event under different treatment conditions. Studying ETEs offers two practical advantages that may facilitate implementation and policy decision making. First, because empirical strata are functions of baseline covariates, stratum membership can be determined once baseline covariates have been collected, thereby enhancing policy relevance. Second, ETEs are identifiable under the standard causal assumptions commonly used in observational causal  frameworks \citep{rosenbaum1983central}, thereby easily facilitating practical implementation. 

We emphasize that 
empirical stratification and principal stratification, while both concerned with PV-related treatment effect heterogeneity, address distinct inferential goals.  Principal stratification asks whether treatment effects vary across latent subgroups defined by counterfactual PV responses. Empirical stratification instead targets treatment-effect variation across baseline-predicted PV response profiles, yielding subgroups that can be identified from baseline covariates and used prospectively for practical decisions. 
Section~\ref{sec:connections} provides a detailed comparison of the two frameworks. Overall, in our paper, principal stratification serves  mainly as a related point of comparison, instead of the research focus. 

\subsection{Our contributions}

Methodologically, this paper makes two main contributions. First, it provides the causal foundation for predictive HTE analysis. 
We define ETEs as a new class of estimands for treatment effect variation across baseline-predicted PV response profiles, and show that they are identifiable under treatment ignorability and positivity assumptions \citep{rosenbaum1983central}. 
These results clarify such empirical subgroup analysis can be interpreted as a valid causal analysis. Second, this paper further develops the semiparametric inference for the efficient estimation of ETEs. The key technical challenge is that the empirical score is not directly observed, but is constructed from estimated baseline predictions of the PV under the two treatment conditions. When baseline covariates are high-dimensional or include some continuous variables, we typically have infinitely many empirical strata, making the estimation of ETEs non-regular and challenging. Building on previous projection-based approaches (e.g., \citealp{lu2026principal,ye2023instrumented,kennedy2019robust,robins2000marginal}), our proposed method projects the ETE onto a prespecified working model to obtain a low-dimensional approximation to the underlying true ETE curve. We refer to the resulting estimand as the projected ETE curve (projETE).  When the working model is correctly specified, the projETE recovers the true ETE. When the working model is misspecified, it remains a well-defined causal summary given by the best least-squares projection of the true ETE onto the chosen working model. We develop a semiparametric estimator of the projETE that combines efficient influence functions (EIFs) with flexible, data-adaptive nuisance estimation \citep{chernozhukov2018double,zheng2011cross}. Together with the cross-fitting algorithm, the proposed  estimator supports parametric-rate inference  when nuisance functions are estimated by off-the-shelf machine learning algorithms with nonparametric rates.

Our method also contributes to the literature on compliance-score subgroup analysis in randomized trials with treatment noncompliance \citep{follmann2000effect,joffe2003compliance,hu2022assessing}. In this setting, prior work studies intention-to-treat effects across strata defined by a compliance score, which summarizes the baseline-predicted probability to comply with treatment assignment. In Remark~\ref{rem:compliance_score}, we clarify that such analyses can be viewed as an important application of our framework by treating treatment receipt as the PV and choosing the empirical score to be the compliance score. Relative to this literature, we extend the compliance-score analysis to a more general empirical stratification framework for PVs beyond treatment noncompliance, and to develop unified EIF-based semiparametric estimation and inference.

\section{The empirical stratification framework}\label{sec:emprical_stratification}

\subsection{Preliminaries and a motivating data example}

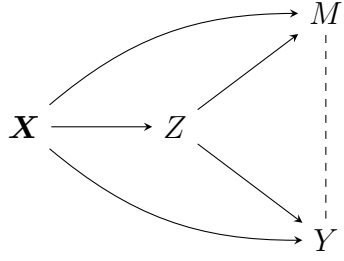
\begin{figure}[t]
\begin{center}
    \begin{tikzpicture}[>=stealth]
  \node (X) at (0,0) {$\bcx$};
  \node (Z) at (2,0) {$Z$};
  \node (M) at (4,1.5) {$M$};
  \node (Y) at (4,-1.5) {$Y$};
  \draw[->] (X) -- (Z);
  \draw[->] (X) to[bend left=20] (M);
  \draw[->] (X) to[bend left=-20] (Y);
  \draw[->] (Z) -- (M);
  \draw[->] (Z) -- (Y);
  \draw[dashed] (M) -- (Y);
\end{tikzpicture}
\end{center}
  \caption{Directed acyclic graph for depicting the causal relationships among $\bcx$, $Z$, $M$, and $Y$. The dashed edge indicates a generic association with an unknown causal structure. In randomized experiments, there is no causal pathway from $\bcx$ to $Z$ due to randomization.}\label{fig:dag}
\end{figure}

Consider that we observe $n$ independent and identically distributed data vectors $\bco_i = \{\bcx_i, Z_i, M_i, Y_i\}$, where $\bcx_i \in \mathcal X \subseteq \mathbb{R}^{p\times 1}$ is a set of baseline covariates, $Z_i \in \{0,1\}$ is a binary treatment (1 for treated; 0 for control), $M_i \in \mathbb{R}$ is a PV, and  $Y_i \in \mathbb{R}$ is an outcome of interest. The $n$ copies of data may be drawn from a randomized trial  or an observational study. We also allow the PV to be either continuously or discretely distributed. Figure~\ref{fig:dag} shows the causal relationship among these variables, where we do not place any restriction on the causal structure between the outcome and PV.  Under the counterfactual outcomes framework, we use $Y_i(z)$ to denote the counterfactual value of outcome when setting $Z_i=z\in\{0,1\}$, and similarly write $M_i(z)$ as the counterfactual value of PV when setting $Z_i=z$. Throughout, we require the Stable Unit Treatment Value Assumption \citep{rubin1986comment} to connect counterfactual and observed values by $Y_i = Y_i(1)Z_i+Y_i(0)\{1-Z_i\}$ and $M_i = M_i(1)Z_i+M_i(0)\{1-Z_i\}$. 
In addition, we assume the following two structural assumptions to identify the proposed estimands:
\begin{assumption}\label{assum:igno}
(Treatment ignorability) $\{Y(1),Y(0), M(1), M(0)\} \indep Z \mid \bcx$.
\end{assumption}
\begin{assumption}\label{assum:posi}
(Positivity) $c <\Prob(Z=1|\bcx)<1-c$ for some $0<c<0.5$.
\end{assumption}
Both assumptions are usually satisfied in randomized trials.  They are the standard set of assumptions for causal inference in observational studies \citep{rosenbaum1983central}. In observational studies, Assumption \ref{assum:igno} requires that, conditional on baseline covariates, treatment assignment behaves ``as if randomized," and Assumption \ref{assum:posi} requires that every individual has a positive probability of receiving each treatment condition. The average treatment effect,  
$
\text{ATE} = \E[Y(1)-Y(0)],
$
is employed to measure the overall treatment effect for the entire study population, but it cannot characterize treatment effect heterogeneity. 

The empirical stratification framework    characterizes the predictive treatment effect heterogeneity with respect to PVs. As a motivating example, we consider the World Health Organization's Large Analysis and Review of European Housing and Health Status (WHO-LARES) study \citep{shenassa2007dampness,cheng2025identification}, which is an observational study of housing conditions among residents in eight European cities.   The effect of residing in damp housing ($Z=1$: yes and $Z=0$: no) on depression ($Y=1$: yes and $Y=0$: no), with some individuals developing dampness-related diseases after treatment ($M=1$: yes and $M=0$: no).  The motivating question is to understand how the effect of damp housing on depression varies with individuals’ dampness-related disease risk. In particular, individuals who are more likely to develop a dampness-related disease may be of special interest for the targeted intervention, because their physical health may be especially vulnerable to damp living conditions. 
To rigorously study this causal effect with such empirical stratification, we next formally develop the corresponding framework.

\subsection{Definition of empirical strata}

Let $\bm G=[M(1),M(0)]^\top$ be the joint counterfactual values of the PV under both treatment conditions, which provides a natural basis for defining PV-induced subgroups. However, $\bm G$ is generally latent because only one of $M(1)$ and $M(0)$ can be observed for each individual. We  use a covariate-based prediction of $\bm G$ to construct empirically accessible subgroups. Specifically, for $z\in\{0,1\}$, let $s_z(\bcx)=\E[M(z)\mid \bcx]$ be the expected PV response under treatment level $z$ based on  covariates $\bcx$. When $M$ is binary, $s_z(\bcx)$ is the baseline  probability (or baseline risk) of the PV under treatment condition $z$. 
When $M$ is continuous, $s_z(\bcx)$ is the baseline-predicted mean level of the PV under treatment condition $z$. Then, the vector
$\E[\bm G \mid \bcx] = [s_1(\bcx), s_0(\bcx)]^\top \in \mathbb{R}^2$
summarizes the predicted PV response profile under both treatment conditions. In practice, investigators may be interested in a particular feature of this predicted response profile rather than the full two-dimensional vector \citep{kent2018personalized,kent2020predictive}. To this end, we define the empirical score as
$$
\tau(\bcx):=r\{s_1(\bcx),s_0(\bcx)\}\in\mathbb R,
$$
where $r(x,y): \mathbb{R}\times\mathbb{R} \rightarrow \mathbb{R}$ is a user-specified transformation function that maps $\{s_1(\bcx), s_0(\bcx)\}$ to a scalar value $\tau(\bcx)$. The empirical score $\tau(\bcx)$ is then used to define empirical strata, namely subgroups formed by values or ranges of $\tau(\bcx)$.  Note that $\tau(\bcx)$ is easily identified based on observed data. Specifically, under Assumptions~\ref{assum:igno}--\ref{assum:posi}, we have $s_z(\bcx) = \E[M(z) \mid \bcx] = \E[M \mid Z=z,\bcx]$, 
so that $\tau(\bcx)$ is identified by
$$
\tau(\bcx)=r\{h_1(\bcx),h_0(\bcx)\} \footnote{For notational convenience, we use the same notation $\tau(\bcx)$ for both the causal definition $r\{s_1(\bcx),s_0(\bcx)\}$ and its identification formula $r\{h_1(\bcx),h_0(\bcx)\}$. Under
Assumptions~\ref{assum:igno}--\ref{assum:posi}, these two quantities are equivalent, so this convention does not create ambiguity.},
$$
where $h_z(\bcx):=\E[M\mid Z=z,\bcx]$. By construction, the choice of the transformation function $r(x,y)$ determines the interpretation of the empirical score. We highlight three example choices and clarify their interpretation in the WHO-LARES study. Web Appendix~1.1 provides three additional data examples illustrating how empirical scores can be interpreted in the applications when the PV denotes a treatment noncompliance status, a behavioral response to treatment, and a secondary outcome. 

\noindent \emph{Choice 1. Treated PV mean.} Choosing $r(x,y)=x$ gives $\tau(\bcx)=s_1(\bcx)$, which stratifies individuals by their predicted PV under treatment. In the WHO-LARES study, $\tau(\bcx)$ represents the probability of developing dampness-related disease under damp housing exposure. Thus, individuals with higher values of $\tau(\bcx)$ are predicted to be more vulnerable to dampness-related disease under exposure.

\noindent\emph{Choice 2. Control PV mean.} Choosing $r(x,y)=y$ gives $\tau(\bcx)=s_0(\bcx)$, which stratifies individuals by their predicted PV under control. In the WHO-LARES study, $\tau(\bcx)$ is the probability of developing dampness-related disease without damp housing exposure, thereby capturing baseline vulnerability to such disease in the absence of exposure.

\noindent\emph{Choice 3. Additive PV contrast.} Choosing $r(x,y)=x-y$ gives $\tau(\bcx)=s_1(\bcx)-s_0(\bcx)$, which stratifies individuals by the predicted treatment-induced change in the PV. In the WHO-LARES study, individuals with higher (or lower) values of $\tau(\bcx)$ are those whose risk of developing dampness-related diseases is predicted to be more (or less) affected by exposure to damp conditions.

\begin{remark}[Alternative definitions of empirical scores]
Here, we define the empirical score using $\E[\bm G\mid \bcx]$, which
summarizes the predicted potential PV responses through their conditional means. More
general definitions of empirical scores are also possible. For example, when $M$ is
continuously distributed, one may define empirical scores using features of the
conditional counterfactual distributions $\Prob[M(1)\leq m_1\mid \bcx]$ and
$\Prob[M(0)\leq m_0\mid \bcx]$, which can capture distributional aspects of the predicted PV
responses beyond their conditional means. In settings where the principal score, $e_{\bm g}(\bcx):=\Prob(\bm G=\bm g\mid \bcx)$, is identifiable, one may also set $\tau(\bcx)=e_{\bm g}(\bcx)$, where $\bm g \in \bm G$ is a fixed principal stratum. In this case, the corresponding ETE has a connection to the principal causal effects; see Section~\ref{sec:connections} for further discussions.
\end{remark}

\subsection{Empirical-stratum treatment effects}

{\color{black}Let $\mathcal{T}:=
    \left\{
        t\in\mathbb R:
        \mathbb{P}[|\tau(\bcx)-t|<\epsilon]>0
        \text{ for every } \epsilon>0
    \right\}$ be the support of the empirical score $\tau(\bcx)$.
} For a fixed empirical stratum $t$ in $\mathcal{T}$, we define the empirical-stratum treatment effect (ETE) as
\begin{equation}\label{eq:ETE}
\mathrm{ETE}(t)
=
\E\{Y(1)-Y(0)\mid \tau(\bcx)=t\}.
\end{equation}
By definition, the ETE characterizes how the treatment effect on the primary outcome varies across individuals with different predicted PV responses, as summarized by $\tau(\bcx)$. The following theorem shows that the ETE is identified based on Assumptions~\ref{assum:igno}--\ref{assum:posi}. 
\begin{theorem}[Identification of the ETE]\label{thm:identification_ETE}
Suppose that Assumptions~\ref{assum:igno}--\ref{assum:posi} hold. Let $\theta_z(t):=\E[Y(z)\mid \tau(\bcx)=t]$ be the counterfactual mean of $Y(z)$ given $\tau(\bcx)=t$ so that $\text{ETE}(t)=\theta_1(t)-\theta_0(t)$. Then, for any $t\in\mathcal{T}$, $\theta_z(t)$ is identified with   
$$
\theta_z(t) =\E\{\mu_z(\bcx)\mid \tau(\bcx)=t\},
$$
where $\mu_z(\bcx)=\E[Y\mid Z=z,\bcx]$. 
\end{theorem}

\begin{figure}[htbp]
    \centering
    \includegraphics[width=0.6\textwidth]{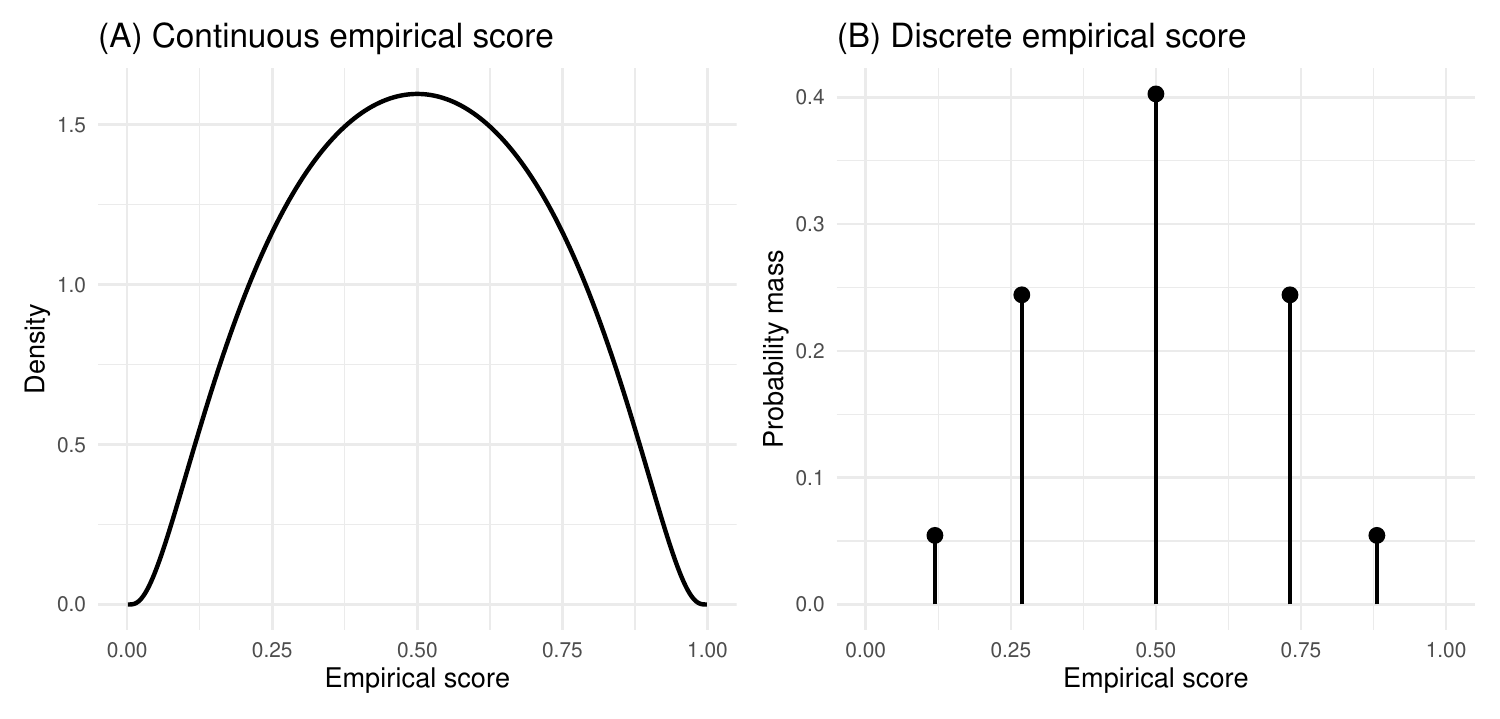}
    \caption{Illustration of two distributional cases for the empirical score $\tau(\bcx)=s_1(\bcx)=\mathrm{expit}(X)$, where $\mathrm{expit}(x)$ is the sigmoid function. In Panel (A), the baseline covariate follows $X\sim N(0,1)$. In Panel (B), $X$ follows a discrete distribution with $\Prob(X=j)\propto \phi(j)$ for $j\in\{-2,-1,0,1,2\}$, where $\phi(x)$ is the standard normal density. By construction, $\tau(\bcx)$ is continuously distributed in Panel (A) and discretely distributed in Panel (B).} 
    \label{fig:density_es_illustration}
\end{figure}

Although Theorem~\ref{thm:identification_ETE} establishes identification of $\mathrm{ETE}(t)$, direct estimation of the ETE pointwisely at $\tau(\bcx)=t$ can be practically challenging. As illustrated in Figure~\ref{fig:density_es_illustration}, the empirical score $\tau(\bcx)$ may be either discrete or continuous.  Both distributional cases indicate estimation challenges for $\text{ETE}(t)$. When $\tau(\bcx)$ is continuous, there are infinitely many empirical strata, making the estimation of   ETE curve a fully nonparametric problem. When $\tau(\bcx)$ is discrete, the number of empirical strata may still be large, especially when the empirical score is constructed from high-dimensional baseline covariates. Some strata may also contain limited probability mass, leading to unstable subgroup-specific estimation. These challenges motivate the projected ETE curves introduced in Section \ref{sec:IF_estimation}, which leverage a parametric working model to provide a low-dimensional summary of the ETE.

\section{The projected ETE curve and its estimation}\label{sec:IF_estimation}

\subsection{Projected ETE curves}

To obtain more   interpretable summaries and stable estimation for $\text{ETE}(t) = \theta_1(t) - \theta_0(t)$, we propose to project each mean curve $\theta_z(t)$ onto a prespecified, low-dimensional parametric working model \citep{robins2000marginal,kennedy2019robust,ye2023instrumented,lu2026principal}. Specifically, for each $z\in\{0,1\}$, let $\eta_z(t;\bm\beta)$ denote a parametric working model for $\theta_z(t)$, which is indexed by a finite, $q$-dimensional parameter $\bm\beta\in\mathbb{R}^{q}$. 
For example, one can specify $\eta_z(t;\bm\beta_z) = \beta_{z,0} + \beta_{z,1}t$ or $\eta_z(t;\bm\beta_z) = \exp(\beta_{z,0} + \beta_{z,1}t)$ with $\bm\beta_z = (\beta_{z,0},\beta_{z,1})^\top$ by modeling $\theta_z(t)$ as a linear or exponential function of $t$. More specifically, we   define the projection parameter $\bm\beta_z$ as
\begin{equation}\label{eq:proj_mz}
\bm\beta_z:=\arg\min_{\bm\beta\in\mathbb{R}^{q}}
\mathbb{E}\!\left[\left\{
\theta_z\bigl(\tau(\bcx)\bigr)-\eta_z\bigl(\tau(\bcx);\bm\beta\bigr)
\right\}^2
\right],
\qquad z\in\{0,1\}.
\end{equation}
Therefore, $\bm\beta_z$ is the least-squares projection of the true mean curve $\theta_z(t)$ onto the model class $\eta_z(t;\bm\cdot)$. When the working model is correctly specified, $\bm\beta_z$ coincides with the model parameter indexing the true mean curve $\theta_z(t)$. When the working model is misspecified,  $\eta_z(t;\bm\beta_z)$ remains a well-defined causal parameter that summarizes the closest approximation to $\theta_z(t)$ based on the least-squares loss function within the chosen model class, namely $\{\eta_z(t,\bm\beta): \bm\beta \in \mathbb{R}^q\}$. Based on projection parameters $\{\bm\beta_0,\bm\beta_1\}$, we define the projected curve for ETE as
\begin{equation*}
\text{projETE}(t) = \eta_1(t;\bm\beta_1)-\eta_0(t;\bm\beta_0), \qquad t\in\mathcal{T},
\end{equation*}
which provides a low-dimensional summary of treatment effect variation as a function of the empirical score. It should be mentioned that when $\eta_z(t;\bm\beta_z)$ has a linear working model with $\eta_z(t;\bm\beta_z) = \beta_{z,0}+\beta_{z,1}t$, the projected ETE curve also has the linear form with
\begin{equation}\label{eq:linear_structural_model}
\mathrm{projETE}(t)
=
\beta_{\mathrm{diff},0}+\beta_{\mathrm{diff},1}t,
\end{equation}
where $\beta_{\mathrm{diff},0}:=\beta_{1,0}-\beta_{0,0}$ and $\beta_{\mathrm{diff},1}:=\beta_{1,1}-\beta_{0,1}$. Here, $\beta_{\mathrm{diff},1}$ summarizes the linear trend in treatment effect heterogeneity along the empirical score; equivalently, a one-unit increase in $\tau(\bcx)$ corresponds to a $\beta_{\mathrm{diff},1}$-unit change in the projected treatment effect. The intercept $\beta_{\mathrm{diff},0}$ represents the projected treatment effect at $\tau(\bcx)=0$, if $\tau(\bcx)=0$ lies in the support of $\tau(\bcx)$. The following proposition shows that $\bm\beta_z$ is identified based on Assumptions~\ref{assum:igno}--\ref{assum:posi}.

\begin{proposition}[Identification of projection parameters]\label{prop:projETE}
Suppose Assumptions~\ref{assum:igno}--\ref{assum:posi} hold. For each $z\in\{0,1\}$, suppose that the projection parameter $\bm\beta_z$ exists and is the unique minimizer of \eqref{eq:proj_mz}. Suppose further that $\eta_z(t;\bm\beta)$ is differentiable in $\bm\beta$ with
{${\eta}_z^{(2)}(t;\bm\beta) := \partial\eta_z(t;\bm\beta)/\partial\bm\beta$}. Then $\bm\beta_z$ is the unique solution of the following equation in terms of $\bm\beta$:
\begin{equation}\label{eq:proj_ee}
\mathbb{E}\!\left[{\eta}_z^{(2)}\bigl(\tau(\bcx);\bm\beta\bigr)
\left\{\mu_z(\bcx)-\eta_z\bigl(\tau(\bcx);\bm\beta\bigr)\right\}\right]=0.
\end{equation}
\end{proposition}

\begin{remark}[
Group-based ETE estimand]\label{remark:grpETE} Our main analysis focuses on the
projected ETE curve, which provides an efficient and stable approximation to
the ETE. Alternatively, one may directly summarize treatment effects across
coarsely defined empirical strata.  In particular,  let
$-\infty=l_0<l_1<\cdots<l_K=+\infty$ define a partition of the support of $\tau(\bcx)$ into $K$ intervals
$\mathcal T_k=[l_{k-1},l_k)$, $k=1,\ldots,K$. In practice, the partition may be chosen based on substantive considerations or empirical quantiles of $\tau(\bcx)$, so that each subgroup has non-negligible probability mass. Then, we can define the group-based ETE as
$$
\mathrm{grpETE}(k) = \E[Y(1)-Y(0)\mid \tau(\bcx)\in\mathcal T_k],
\qquad k=1,\ldots,K.
$$
This estimand compares treatment effects across subgroups defined by the ranges of predicted PV responses. Under Assumptions~\ref{assum:igno}--\ref{assum:posi}, and provided $\Prob\{\tau(\bcx)\in\mathcal T_k\}>0$, the group-based ETE can be identified as
$
\mathrm{grpETE}(k)
=
{
\E\!\left[
\mathbb{I}\{\tau(\bcx)\in\mathcal T_k\}
\{\mu_1(\bcx)-\mu_0(\bcx)\}
\right]
}\Big/{
\Prob\{\tau(\bcx)\in\mathcal T_k\}
}.
$
Detailed inference procedures for group-based ETEs, including the EIF-based estimator and its asymptotic behavior, are provided in Web Appendix 1.2 of the Supplementary Material.
\end{remark}

\subsection{Moment-type estimator}

We first propose a moment-type estimator for the projection parameter $\bm\beta_z$, which is motivated by the identification formula in Proposition \ref{prop:projETE}.  For $z\in\{0,1\}$, let $\widehat h_z(\bcx)$ and $\widehat \mu_z(\bcx)$ be estimators of the two nuisance parameters $h_z(\bcx)=\E[M\mid Z=z,\bcx]$ and $\mu_z(\bcx)=\E[Y\mid Z=z,\bcx]$, which can be obtained through regressing $Y$ and $M$ on $\{Z,\bcx\}$. Based on $\widehat h_z(\bcx)$, $\tau(\bcx)$ can be immediately obtained by $\widehat\tau(\bcx)= r\{\widehat h_1(\bcx),\widehat h_0(\bcx)\}$. Motivated by Proposition \ref{prop:projETE}, the moment-type estimator of $\bm\beta_z$ for $z\in\{0,1\}$, denoted by $\widehat{\bm\beta}_z^{\mathrm{mo}}$, can be obtained by solving the following estimating equation in terms of $\bm\beta$
$$
\frac{1}{n}\sum_{i=1}^n\left[\eta_z^{(2)}\bigl(\widehat \tau(\bcx_i);\bm\beta\bigr)
\left\{\widehat \mu_{z}(\bcx_i)-\eta_z\bigl(\widehat\tau(\bcx_i);\bm\beta\bigr)\right\}\right]=0,
$$
where we replace the population-level expectation operator ``$\E$" in the identification formula with the empirical average operator ``$\frac{1}{n}\sum_{i=1}^n$" across all data points, and also replace the unknown nuisance parameters $\{\tau(\bcx),\mu_{z}(\bcx)\}$ by their estimates $\{\widehat \tau(\bcx),\widehat \mu_{z}(\bcx)\}$. After obtaining $\widehat{\bm\beta}_z^{\mathrm{mo}}$, the projected curve $\text{projETE}(t)$ can be estimated by 
$$
\widehat{\text{projETE}}^{\mathrm{mo}}(t) =  \eta_1(t;\widehat{\bm\beta}_1^{\mathrm{mo}})-\eta_0(t;\widehat{\bm\beta}_0^{\mathrm{mo}}).
$$

Although $\widehat{\bm\beta}_z^{\mathrm{mo}}$ is simple to implement, its large-sample behavior is sensitive to first-stage estimation error in the nuisance parameters, $\{\widehat \tau(\bcx),\widehat \mu_{z}(\bcx)\}$. Therefore, $\widehat{\bm\beta}^{\mathrm{mo}}$ may be neither $\sqrt{n}$-consistent nor efficient. This motivates the development of EIF-based estimators in Section \ref{sec:eifPE}--\ref{sec:EIF_estimator}.
We develop EIF-based estimators for the projected ETE curve using semiparametric theory \citep{chernozhukov2018double,zheng2011cross}. In contrast to the moment-type estimator, the EIF-based estimators reduce sensitivity to first-stage nuisance estimation error and support  parametric-rate inference when nuisance parameters, such as $\mu_z(\bcx)$ and $h_z(\bcx)$, are estimated at nonparametric rates. 

\subsection{The EIF of the projection parameter}\label{sec:eifPE}

We first derive the EIF of the projection parameter $\bm\beta_z$ for $z\in\{0,1\}$ under the nonparametric model of the observed data $\bco_i$. Following the semiparametric theory  \citep{bickel1993efficient,hines2022demystifying}, to characterize the EIF, we define a parametric submodel $\mathbb{P}_v=(1-v)\Prob + v\widetilde{\Prob}$ for the distribution of the observed data $\bco_i$, where $v\in[0,1]$ is a one-dimensional parameter indexing the submodel, $\Prob$ is the true distribution of $\bco_i$, and $\widetilde \Prob \neq \Prob$ is another distribution of $\bco_i$. Let $\bm\beta_z(\Prob_v)$ be the value of $\bm\beta_z$ evaluated under the parametric submodel $\Prob_v$, which is characterized as the solution of 
$$
\E_{\bm X\sim\mathbb{P}_v}
\left[\eta_z^{(2)}\bigl(\tau(\bm X;\Prob_v);\bm\beta\bigr)
\left\{\mu_z(\bm X;\Prob_v)-\eta_z\bigl(\tau(\bm X;\Prob_v);\bm\beta\bigr)\right\} \right]=0
$$
with respect to $\bm\beta\in \mathbb R^q$, where $\{\tau(\bcx;\Prob_v),\mu_z(\bcx;\Prob_v)\}$ are the values of $\{\tau(\bcx),\mu_z(\bcx)\}$   under $\Prob_v$. The true value $\bm\beta_z$ is therefore $\bm\beta_z(\Prob_0)$. The EIF of $\bm\beta_z$, if exists, is the unique mean-zero, finite variance function $\psi_{\bm\beta_z}^{\mathrm{eff}}(\bco)$ with $\frac{d}{d v} \bm\beta_z(\Prob_v) \Big|_{v=0} = \int \psi_{\bm\beta_z}^{\mathrm{eff}}(\bm o_i) d \widetilde{\Prob}(\bm o_i)$, where $\frac{d}{d v}$ is the pathwise (or directional) derivative with respect to $v$. Therefore, a standard condition for deriving the EIF is to require the pathwise derivative of the target estimand $\bm\beta_z(\Prob_v)$ with respect to $v$ is well defined. To ensure that the EIF exists, we introduce the following regularity condition requiring that the working model, $\eta_z(t,\bm\beta)$, and the transformation function for defining the empirical score, $r(x,y)$, are both sufficiently smooth.
\begin{assumption}(Smoothness)\label{assum:smoothness}
Let $\mathcal N(\bm\beta_z)=\{\bm\beta\in\mathbb R^q:\|\bm\beta-\bm\beta_z\|\leq\delta\}$ be a fixed compact neighborhood of $\bm\beta_z$ for some constant $\delta>0$. Suppose the following conditions hold.
\begin{compactitem}
\item[(i)] For both $z\in\{0,1\}$,  $\eta_z(t;\bm\beta)$ is twice continuously differentiable, where $\eta_z^{(1)}(t;\bm\beta)=\partial\eta_z(t;\bm\beta)/\partial t$, $\eta_z^{(2)}(t;\bm\beta)=\partial\eta_z(t;\bm\beta)/\partial\bm\beta$, $\eta_z^{(21)}(t;\bm\beta)=\partial^2\eta_z(t;\bm\beta)/\partial\bm\beta\partial t$, and $\eta_z^{(22)}(t;\bm\beta)=\partial^2\eta_z(t;\bm\beta)/\partial\bm\beta\partial\bm\beta^\top$ are uniformly bounded and  Lipschitz on $(t,\bm\beta) \in \mathcal T\times\mathcal N(\bm\beta_z)$.

\item[(ii)] $r(x,y)$ is differentiable, where $r^{(1)}(x,y)=\partial r(x,y)/\partial x$ and $r^{(2)}(x,y)=\partial r(x,y)/\partial y$ are uniformly bounded and Lipschitz on $(x,y)\in \mathcal H_1\times \mathcal H_0$. Here, $\mathcal H_z$ is the support of $h_z(\bcx)$.
\end{compactitem}
\end{assumption}

Under this smoothness condition, we derive the explicit expression for the EIF of $\bm\beta_z$. 
\begin{proposition}\label{prop:proj_eif}
Suppose that Assumptions \ref{assum:igno}--\ref{assum:smoothness} hold. In addition, assume that the following $q$-by-$q$ negative Jacobian matrix of the estimating function \eqref{eq:proj_ee} exists and is nonsingular:
$$
H_z(\bm\beta_z)
=\E\left[\left\{\eta_z^{(2)}(\tau(\bcx),\bm\beta_z)\right\}^{\otimes 2}-
    \eta_z^{(22)}(\tau(\bcx),\bm\beta_z)
    \left\{\mu_z(\bcx)-\eta_z(\tau(\bcx),\bm\beta_z)\right\}
    \right],$$
where $\bm b^{\otimes 2} = \bm b \bm b^\top$ denotes the outer product for some vector $\bm b$. Define $\pi_z(\bcx)=\Prob(Z=z|\bcx)$ as the treatment probability, the EIF of $\bm\beta_z$ exists with the following expression:
$$
\psi_{\bm\beta_z}^{\mathrm{eff}}(\bco) = \left\{H_z(\bm\beta_z)\right\}^{-1} \Omega_z\left(\bco;\bm\beta_z,\mu_z,h_1,h_0,\pi_1,\pi_0\right),
$$
where $\Omega_z=\Omega_z\left(\bco;\bm\beta_z,\mu_z,h_1,h_0,\pi_1,\pi_0\right)$ is a $q$-by-$1$ dimensional function of the data $\bco$, the projection parameter $\bm\beta_z$, and nuisance parameters $\{\mu_z,h_1,h_0,\pi_1,\pi_0\}$:
\begin{align*}
\Omega_z = &  \eta_z^{(2)}(\tau(\bcx),\bm\beta_z)\left[\mu_z(\bcx)-\eta_z(\tau(\bcx),\bm\beta_z)\right] \\
&  + \eta_z^{(2)}(\tau(\bcx),\bm\beta_z)\frac{\mathbb{I}(Z=z)}{\pi_z(\bcx)}\left\{Y-\mu_z(\bcx)\right\} \\
&  + \Big[\eta_z^{(21)}(\tau(\bcx),\bm\beta_z)\left\{\mu_z(\bcx)-\eta_z(\tau(\bcx),\bm\beta_z)\right\}-\eta_z^{(2)}(\tau(\bcx),\bm\beta_z) \eta_z^{(1)}(\tau(\bcx),\bm\beta_z)\Big] \times \\
&  \quad \left[\frac{Z r^{(1)}(h_1(\bcx), h_0(\bcx))}{\pi_1(\bcx)} \left\{M-h_1(\bcx)\right\} + \frac{(1-Z) r^{(2)}(h_1(\bcx), h_0(\bcx))}{\pi_0(\bcx)} \left\{M-h_0(\bcx)\right\} \right].
\end{align*}
As a result, the nonparametric efficiency bound for estimating $\bm\beta_z$ is $\E\left[\left\{\psi_{\bm\beta_z}^{\mathrm{eff}}(\bm O)\right\}^{\otimes 2}\right]$.
\end{proposition}

The EIF in Proposition~\ref{prop:proj_eif} consists of the inverse Jacobian matrix $H_z(\bm\beta_z)^{-1}$ multiplied by the estimating function  $\Omega_z(\bco;\bm\beta_z,\mu_z,h_1,h_0,\pi_1,\pi_0)$, and the latter has three terms. The first term is the original estimating function in the identification formula \eqref{eq:proj_ee}. The second term augments this estimating function with the outcome residual $Y-\mu_z(\bcx)$, weighted by the inverse treatment probability $\pi_z(\bcx)^{-1}$ and the derivative of working model $\eta_z^{(2)}(\tau(\bcx),\bm\beta_z)$. The third term is a weighted PV-residual term involving $M-h_1(\bcx)$ and $M-h_0(\bcx)$ to correct for the first-order impact of estimating the empirical score $\tau(\bcx)=r\{h_1(\bcx),h_0(\bcx)\}$. Thus, although the EIF has a somewhat involved expression, it depends only on observable data, the projection parameter $\bm\beta_z$, and the nuisance parameters $\{\mu_z,h_1,h_0,\pi_1,\pi_0\}$.

\subsection{EIF-based estimator and its asymptotic property}\label{sec:EIF_estimator}

Because the expectation of EIF is zero, Proposition \ref{prop:proj_eif} motivates a new estimator of $\bm\beta_z$ by solving the following EIF-based estimating equation with respect to $\bm\beta$, if the nuisance parameters $\{\mu_z,h_1,h_0,\pi_1,\pi_0\}$ were known:
$$
\frac{1}{n} \sum_{i=1}^n \left\{H_z(\bm\beta_z)\right\}^{-1} \Omega_z\left(\bco_i;\bm\beta,\mu_z,h_1,h_0,\pi_1,\pi_0\right) = 0 
$$
where $H_z(\bm\beta_z)$ is a constant matrix that does not affect the solution of the previous estimating equation. In practice, the nuisance parameters are unknown and need to be estimated at first. In this paper, we mainly consider using data-adaptive or machine learning-based approaches (e.g., random forest \citep{breiman2001random}, the gradient boosting machine \citep{friedman2001greedy}) to obtain the nuisance parameters, $\pi_z(\bcx)=\Prob(Z=z\mid \bcx)$, $h_z(\bcx)=\E[M \mid Z=z,\bcx]$, and $\mu_z(\bcx)=\E[Y\mid Z=z,\bcx]$.  In this setting, it is well known that directly plugging in the nuisance parameter estimates into the previous EIF-based estimating equation may suffer from overfitting bias arising from using the same data to estimate nuisance parameters and evaluate the target estimand  \citep{chernozhukov2018double}. We propose using the cross-fitting procedure to mitigate the overfitting bias, which leads to the following procedure to calculate the EIF-based estimator, $\widehat{\bm\beta}_z^{\mathrm{eff}}$:
\begin{compactitem}
\item[1.] Take an $L\geq 2$ fold random partition $\{\mathcal D_1,\cdots, \mathcal D_L\}$ of data indices $\{1,\dots,n\}$ such that each fold has approximately the same size. For each data fold $\mathcal D_l$, obtain the nuisance parameter estimates $\{\widehat \mu_{z,l}(\bcx_i),\widehat h_{1,l}(\bcx_i),\widehat h_{0,l}(\bcx_i),\widehat\pi_{1,l}(\bcx_i),\widehat\pi_{0,l}(\bcx_i)\}$ for all $i\in \mathcal D_l$ based on machine learners trained on data $\mathcal D_{-l} = \{1,\dots,n\}\backslash \mathcal D_l$.
\item[2.] Then, $\widehat{\bm\beta}_z^{\mathrm{eff}}$ is obtained by solving the cross-fitted EIF-based estimating equation
$$
\frac{1}{n} \sum_{l=1}^L\sum_{i\in \mathcal D_l} \Omega_z\left(\bco_i;\bm\beta,\widehat \mu_{z,l},\widehat h_{1,l},\widehat h_{0,l}, \widehat \pi_{1,l},\widehat \pi_{0,l}\right) = 0 
$$
\item[3.] The estimated variance of $\widehat{\bm\beta}_z^{\mathrm{eff}}$ (scaled by $\sqrt{n}$), denoted by $\widehat\Sigma_{\bm\beta_z}^{\mathrm{eff}}$, is 
$$
\widehat\Sigma_{\bm\beta_z}^{\mathrm{eff}} = \frac{1}{n} \sum_{l=1}^L\sum_{i\in \mathcal D_l} \left\{\widehat H_z(\widehat{\bm\beta}_z^{\mathrm{eff}})\right\}^{-1} \left\{\Omega_z\left(\bco_i;\widehat{\bm\beta}_z^{\mathrm{eff}},\widehat \mu_{z,l},\widehat h_{1,l},\widehat h_{0,l}, \widehat \pi_{1,l},\widehat \pi_{0,l}\right)\right\}^{\otimes 2} \left\{\widehat H_z(\widehat{\bm\beta}_z^{\mathrm{eff}})\right\}^{-\top}
$$
where
$$
\widehat H_z(\widehat{\bm\beta}_z^{\mathrm{eff}}) = \frac{1}{n} \sum_{l=1}^L\sum_{i\in \mathcal D_l} \left\{\eta_z^{(2)}(\widehat\tau_l(\bcx),\widehat{\bm\beta}_z^{\mathrm{eff}})\right\}^{\otimes 2}  \!- \eta_z^{(22)}(\widehat \tau_l(\bcx),\widehat{\bm\beta}_z^{\mathrm{eff}}) \{\widehat \mu_{z,l}(\bcx)-\eta_z(\widehat\tau_l(\bcx),\widehat{\bm\beta}_z^{\mathrm{eff}})\}.
$$
\end{compactitem}

\begin{theorem}\label{thm:asymototic_beta}
 Let $\gamma = \{\mu_z,h_1,h_0,\pi_1,\pi_0\}$ denote all nuisance parameters. Suppose  Assumptions \ref{assum:igno}--\ref{assum:smoothness} with the following regularity conditions hold for any $l\in\{1,\dots,L\}$.
\begin{compactitem}
\item[(i)] With $\widehat\gamma_l$ fixed, the function classes $\{\Omega_z(\bm o;\bm\beta,\gamma):\bm\beta\in\mathcal N(\bm\beta_z)\}$, $\{\Omega_z(\bm o;\bm\beta,\widehat\gamma_l):\bm\beta\in\mathcal N(\bm\beta_z)\}$, $\{\partial\Omega_z(\bm o;\bm\beta,\gamma)/\partial\bm\beta^\top:\bm\beta\in\mathcal N(\bm\beta_z)\}$, and $\{\partial\Omega_z(\bm o;\bm\beta,\widehat\gamma_l)/\partial\bm\beta^\top:\bm\beta\in\mathcal N(\bm\beta_z)\}$ are $\Prob$-Donsker with square-integrable envelope functions. In addition, the Jacobian matrix $H_z(\bm\beta)$ is nonsingular for all $\bm\beta\in\mathcal N(\bm\beta_z)$. 
\item[(ii)] There exists $b_1\in(0,0.5)$ and $b_2>0$ such that $b_1<\pi_{z}(\bm x)<1-b_1$, $b_1<\widehat \pi_{z,l}(\bm x)<1-b_1$, $|\mu_{z}(\bm x)|<b_2$, $|\widehat \mu_{z,l}(\bm x)|<b_2$, $|h_{z}(\bm x)|<b_2$, $|\widehat h_{z,l}(\bm x)|<b_2$ for any $\bm x \in \mathcal X$ and $z\in\{0,1\}$. 
\item[(iii)] Assume the nuisance parameters are consistently estimated with $\|\widehat \pi_{z,l}(\bcx) - \pi_z(\bcx)\|_2=o_\Prob(1)$, $\|\widehat h_{z,l}(\bcx) - h_z(\bcx)\|_2=o_\Prob(1)$, and $\|\widehat \mu_{z,l}(\bcx) - \mu_z(\bcx)\|_2 = o_\Prob(1)$ for both $z\in\{0,1\}$, where $\|\cdot\|_q$ denotes the $L_q(\Prob)$-norm. Furthermore, the convergence rates of the nuisance parameter estimates satisfy:
\begin{align*}
& \|\widehat \pi_{z,l}(\bcx)-\pi_z(\bcx)\|_2\times \|\widehat \mu_{z,l}(\bcx)-\mu_z(\bcx)\|_2 = o_\Prob(n^{-1/2}),\\
& \|\widehat \pi_{z,l}(\bcx)-\pi_z(\bcx)\|_2\times \|\widehat h_{z,l}(\bcx)-h_z(\bcx)\|_2 = o_\Prob(n^{-1/2}),\\
& \|\widehat h_{z,l}(\bcx)-h_z(\bcx)\|_2 = o_\Prob(n^{-1/4}).
\end{align*}
\end{compactitem}
Then, 
$
\sqrt{n} \left\{\widehat{\bm\beta}_z^{\mathrm{eff}} - \bm\beta_z\right\} \xrightarrow{d} N\left(0,\Sigma_{\bm\beta_z}^{\mathrm{eff}}\right),
$
where $\Sigma_{\bm\beta_z}^{\mathrm{eff}} = \E\left[\left\{\psi_{\bm\beta_z}^{\mathrm{eff}}(\bm O)\right\}^{\otimes 2}\right]$ is the variance of the EIF. In addition, the proposed variance estimator is consistent with $\widehat\Sigma_{\bm\beta_z}^{\mathrm{eff}} = \Sigma_{\bm\beta_z}^{\mathrm{eff}} + o_\Prob(1)$. 
\end{theorem}

Theorem \ref{thm:asymototic_beta} develops the asymptotic behavior of $\widehat{\bm\beta}_z^{\mathrm{eff}}$, which requires three regularity conditions. First, condition (i) imposes a stochastic equicontinuity condition on the EIF-based estimating functions and their derivatives. With the nuisance estimators fixed at $\widehat\gamma_l$, these function classes indexed by $\bm\beta\in\mathcal N(\bm\beta_z)$ are required to be $\Prob$-Donsker. Notably, this condition only ensures stochastic equicontinuity with respect to $\bm\beta$, and places no restriction on the complexity of the nuisance parameter estimators. Condition (i) also requires that the Jacobian matrix is invertible. Condition (ii) places boundedness conditions on the nuisance parameters and their estimates. Condition (iii) requires that all nuisance parameters are consistently estimated under the $L_2(\Prob)$ norm, which further characterizes the convergence rate conditions.  Specifically, if all nuisance parameters $\{\pi_1,\pi_0,h_1,h_0,\mu_z\}$ are estimated at an $o_\Prob(n^{-1/4})$ rate, then the EIF-based estimator of the projection parameter, $\widehat{\bm\beta}^{\mathrm{eff}}_z$, is $\sqrt n$-consistent and asymptotically normal. Moreover, the consistency of the accompanying variance estimator justifies a standard Wald-type confidence interval for $\bm\beta_z$.

\subsection{Inference on the projected ETE curve}

For any fixed $t\in\mathcal{T}$, the projected ETE curve  can be estimated by $\widehat{\mathrm{projETE}}^{\mathrm{eff}}(t)=\eta_1(t;\widehat{\bm\beta}^{\mathrm{eff}}_1)-\eta_0(t;\widehat{\bm\beta}^{\mathrm{eff}}_0)$.  Corollary \ref{cor:projETE} establishes the pointwise asymptotic behavior of $\widehat{\mathrm{projETE}}^{\mathrm{eff}}(t)$:
\begin{corollary}[Asymptotic behavior for the projected ETE curve]
\label{cor:projETE}
Suppose that the conditions in Theorem~\ref{thm:asymototic_beta} hold for both $\bm\beta_1$ and $\bm\beta_0$. For any fixed $t\in\mathcal{T}$, we have that 
$$
\sqrt{n}\left\{
\widehat{\mathrm{projETE}}^{\mathrm{eff}}(t)
-
\mathrm{projETE}(t)
\right\}
\overset{d}{\longrightarrow}
N\{0,\sigma^2_{\mathrm{proj}}(t)\},
$$
where $\sigma^2_{\mathrm{proj}}(t)=\bm a(t)^\top\Sigma_{\bm\beta}\bm a(t)$, $\bm a(t)=\begin{pmatrix}
\eta_1^{(2)}(t;\bm\beta_1) \\
-\eta_0^{(2)}(t;\bm\beta_0)
\end{pmatrix}$, and 
$
\Sigma_{\bm\beta}
=
\E\left[
\begin{pmatrix}
\psi_{\bm\beta_1}^{\mathrm{eff}}(\bm O)\\
\psi_{\bm\beta_0}^{\mathrm{eff}}(\bm O)
\end{pmatrix}^{\otimes 2}
\right]
$
is the joint asymptotic covariance matrix of $\widehat{\bm\beta}^{\mathrm{eff}}_1$ and $\widehat{\bm\beta}^{\mathrm{eff}}_0$. 
\end{corollary}

For purposes of inference, we need to estimate the asymptotic variance $\sigma^2_{\mathrm{proj}}(t)$. Let $\widehat{\psi}_{\bm\beta}^{\mathrm{eff}}(\bco_i)=\{(\widehat\psi_{\bm\beta_1}^{\mathrm{eff}}(\bm O_i))^\top,(\widehat\psi_{\bm\beta_0}^{\mathrm{eff}}(\bm O_i))^\top\}^\top$, where
$$
\widehat\psi_{\bm\beta_z}^{\mathrm{eff}}(\bm O_i) =  \left\{\widehat H_z(\widehat{\bm\beta}_z^{\mathrm{eff}})\right\}^{-1} \left\{\Omega_z\left(\bco_i;\widehat{\bm\beta}_z^{\mathrm{eff}},\widehat \mu_{z,l},\widehat h_{1,l},\widehat h_{0,l}, \widehat \pi_{1,l},\widehat \pi_{0,l}\right)\right\}
$$
is the cross-fitted plug-in estimate of the EIF of $\bm\beta_z$. Then, $\sigma^2_{\mathrm{proj}}(t)$ can be estimated by $\widehat{\sigma}^2_{\mathrm{proj}}(t)
= \widehat{\bm a}(t)^\top \widehat{\Sigma}_{\bm\beta} \widehat{\bm a}(t)$ with $\widehat{\bm a}(t)=\left[\{\eta_1^{(2)}(t;\widehat{\bm\beta}^{\mathrm{eff}}_1)\}^\top,-\{\eta_0^{(2)}(t;\widehat{\bm\beta}^{\mathrm{eff}}_0)\}^\top\right]^\top$ and $\widehat{\Sigma}_{\bm\beta} = \frac{1}{n} \sum_{l=1}^L\sum_{i\in \mathcal D_l} \left\{\widehat{\psi}_{\bm\beta}^{\mathrm{eff}}(\bco_i)\right\}^{\otimes 2}$. Finally, for a fixed $t\in \mathcal T$, a $(1-\alpha)$ confidence interval for $\mathrm{projETE}(t)$ is given by $\widehat{\mathrm{projETE}}^{\mathrm{eff}}(t)\pm z_{1-\alpha/2}\sqrt{\widehat{\sigma}^2_{\mathrm{proj}}(t)/n}$, where $z_{1-\alpha/2}$ is the $(1-\alpha/2)$ lower quantile of the standard normal distribution.

\section{Simulation studies}\label{sec:simulation}

\subsection{Set-up}

We conduct simulation studies to evaluate finite-sample performance of the proposed estimators. We simulate $1,000$ observational studies each with $n=1,000$ individuals. For individual $i\in\{1,\dots,n\}$, we first generate the full data $\bm W_i = \{\bcx_i,Z_i,M_i(1),M_i(0),Y_i(1),Y_i(0)\}$ as follows, and the observed data $\bco_i = \{\bcx_i,Z_i,M_i = M_i(Z_i),Y_i = Y_i(Z_i)\}$ can be extracted from $\bm W_i$. Specifically, we first generate $\bm X_i = [X_{i1},X_{i2},X_{i3}]^\top$ based on three independent, standard normal distributions. Then, $Z_i$ is generated from a Bernoulli distribution with $\Prob(Z_i=1\mid \bcx_i) = \text{expit}(0.3X_{i1}+0.3X_{i2}+0.3X_{i3})$, where $\text{expit}(x)=\exp(x)/\{1+\exp(x)\}$. Next, we consider a binary PV with $\{M_i(1),M_i(0)\} \in \{0,1\}^{\otimes 2}$ sampled from the following Gaussian copula model \citep{masarotto2012gaussian}:
\begin{align*}
\Prob\left\{M_i(1)\leq m_1, M_i(0)\leq m_0\mid \bcx_i\right\} = \Phi_2 \left\{\Phi^{-1}\left(\Pi_1(m_1)\right),\Phi^{-1}\left(\Pi_0(m_0)\right)\right\},
\end{align*}
where marginally $\Pi_z(m_z) = \Prob(M_i(z)\leq m_z\mid \bcx_i)$ follows a logistic model with $\Prob(M_i(z)=1|\bcx_i) = \text{expit}(-0.75 + 1.5 z + 0.3X_{i1}+0.3X_{i2}+0.3X_{i3})$, $\Phi^{-1}(\cdot)$ is the inverse CDF of standard normal distribution, and $\Phi_2(\cdot, \cdot)$ is the CDF of the bivariate standard normal distribution with correlation 0.5. Similarly, conditioning on $\{M_i(1),M_i(0),\bcx_i\}$, we generate a continuous outcome with $\{Y_i(1),Y_i(0)\} \in \mathbb{R}^2$ sampled from the Gaussian copula model
\begin{align*}
\Prob\left\{Y_i(1)\leq y_1, Y_i(0)\leq y_0\mid M_i(1),M_i(0),\bcx_i\right\} = \Phi_2 \left\{\Phi^{-1}\left(\mathcal Y_1(y_1)\right),\Phi^{-1}\left(\mathcal Y_0(y_0)\right)\right\},
\end{align*}
where marginally $\mathcal Y_z(y_z) = \Prob(Y_i(z)\leq y_z\mid M_i(1),M_i(0),\bcx_i)$ follows a normal distribution $N\Big(2 + z(3+ 2X_{i1} + 0.5X_{i2} + 0.5X_{i3}) +  (3+z+2X_{i1})M(z) + 2X_{i2} + 2X_{i3},4^2\Big)$. We define the empirical score as $\tau(\bcx) = s_1(\bcx)$ to 
  study the projected ETE curve, $\text{projETE}(t) = \eta_1(t,\bm\beta_1) - \eta_0(t,\bm\beta_0)$ with linear working models of $\eta_1(t,\bm\beta_1) = \beta_{1,0} + \beta_{1,1}t$ and $\eta_0(t,\bm\beta_0) = \beta_{0,0} + \beta_{0,1}t$. Based on such linear working models, the projected curve of ETE is $\text{projETE}(t) = \beta_{\mathrm{diff},0}+\beta_{\mathrm{diff},1}t$ in \eqref{eq:linear_structural_model}. The true values of the projection parameters $\bm\beta_z$ (for $z=0,1$) are calculated based on \eqref{eq:proj_mz} from a simulated superpopulation with $n=1,000,000$ individuals assuming that the full data $\bm W_i$ is known. 

For any projection parameter $\Delta \in \{\beta_{1,0},\beta_{1,1},\beta_{0,0},\beta_{0,1},\beta_{\mathrm{diff},0},\beta_{\mathrm{diff},1}\}$, we compare the relative bias, Monte Carlo standard deviation, and 95\% confidence interval coverage across the following three estimators: (i) the moment-type estimator ($\widehat \Delta^{\mathrm{mo}}$) using parametric models for nuisance parameters; (ii) the EIF-based estimator ($\widehat \Delta^{\mathrm{eff-par}}$) using parametric models for nuisance parameters, (iii) the EIF-based estimator ($\widehat \Delta^{\mathrm{eff-ml}}$) using machine learning algorithms for nuisance parameters. When parametric models are considered for estimating nuisance parameters $\{\pi_z(\bcx),h_z(\bcx),\mu_z(\bcx)\}$, the treatment probability $\pi_z(\bcx)$ is modeled by a logistic regression of $Z_i$ on $\bcx_i$, the conditional expectation of PV $h_z(\bcx)$ is obtained by fitting a logistic model by regressing $M_i$ on $\bcx_i$ among the subset of individuals with $Z_i=z$, the outcome expectation $\mu_z(\bcx)$ is modeled by a linear model of $Y_i$ on $\bcx_i$ among subset individuals with $Z_i=z$, respectively. When machine learning algorithms are considered, we consider a 10-fold cross-fitting algorithm, where $\pi_z(\bcx)$ is estimated by regressing $Z_i$ on $\bcx_i$ based on the \texttt{SuperLearner} package in \texttt{R} software  incorporating the generalized linear model and random
forest libraries \citep{phillips2023practical}. Similarly, $h_z(\bcx)$ (respectively, $\mu_z(\bcx)$) is obtained by regressing $M_i$ (respectively, $Y_i$) on $\bcx_i$ among subset individuals with $Z_i=z$ based on \texttt{SuperLearner} with the same libraries.  Although machine learning methods are more flexible than parametric models for nuisance estimation, their finite-sample performance can still depend on the choice of input features. We therefore consider two simulation scenarios to assess the sensitivity of the proposed estimators to the feature representation used in the nuisance models. In Scenario~A, we use the original covariates $\bcx_i=[X_{i1},X_{i2},X_{i3}]^\top$ as inputs for both the parametric and machine learning nuisance estimators. This represents a relatively favorable setting for nuisance estimation. In Scenario~B, we create a more challenging nuisance-estimation setting by constructing three transformed covariates, $U_{i1}=\exp(0.5X_{i1})$, $U_{i2}=\exp(0.5X_{i2})$, and $U_{i3}=\sin(X_{i3})$, and replacing $\bcx_i$ in the nuisance-model feature matrix with $[U_{i1},U_{i2},U_{i3}]^\top$. 

\subsection{Simulation results}

Table~\ref{tab:sim_proj_1000} summarizes the finite-sample performance of the proposed estimators. Under Scenario A, where nuisance estimation is relatively well behaved because the original covariates are used in fitting the parametric and machine learning models, all three estimators perform reasonably well, with small relative biases and coverage probabilities close to the nominal 95\% level. The EIF-based estimators, $\widehat\Delta^{\mathrm{eff-par}}$ and $\widehat\Delta^{\mathrm{eff-ml}}$, have slightly larger Monte Carlo standard deviations than the moment-type estimator $\widehat\Delta^{\mathrm{mo}}$. Under Scenario B, where nuisance estimation is more challenging because the fitted models use transformed covariates, the advantage of the EIF-based estimators becomes more evident. The moment-type estimator performs poorly for $\{\beta_{1,0},\beta_{1,1},\beta_{\mathrm{diff},0},\beta_{\mathrm{diff},1}\}$, exhibiting substantial bias and severe undercoverage, although its performance for $\{\beta_{0,0},\beta_{0,1}\}$ remains relatively stable. In contrast, both EIF-based estimators substantially reduce bias and improve coverage, with coverage probabilities above 93\% for all projection parameters in Scenario B. Overall, these results suggest that the EIF-based estimators are more robust to first-stage nuisance estimation error than the moment-type estimator, particularly when nuisance parameters are more difficult to estimate. 

\renewcommand{\arraystretch}{1.4}
\begin{table}[t]
\centering
\caption{Simulation results for the projection parameters with $n=1,000$ individuals.\label{tab:sim_proj_1000}}

\newcommand{\tablewidth}{1.00\linewidth}

\vspace{-0.4cm}

\resizebox{\tablewidth}{!}{%
\fontsize{11}{11}\selectfont
\begin{tabular}{lccccccccc}
\toprule
\multicolumn{1}{l}{ } & \multicolumn{3}{c}{Moment-type ($\widehat \Delta^{\mathrm{mo}}$)} & \multicolumn{3}{c}{EIF-based ($\widehat \Delta^{\mathrm{eff-par}}$)} & \multicolumn{3}{c}{EIF-based ($\widehat \Delta^{\mathrm{eff-ml}}$)} \\
\cmidrule(l{3pt}r{3pt}){2-4}
\cmidrule(l{3pt}r{3pt}){5-7}
\cmidrule(l{3pt}r{3pt}){8-10}
Estimand ($\Delta$) & RelBias & MCSD & Coverage & RelBias & MCSD & Coverage & RelBias & MCSD & Coverage\\
\midrule
\addlinespace[0.3em]
\multicolumn{10}{l}{\textbf{Scenario A: original covariates}}\\
\hspace{1em}$\beta_{1,0}$ & $-0.005$ & $1.049$ & $0.950$ & $0.064$ & $1.347$ & $0.956$ & $0.075$ & $1.404$ & $0.952$\\
\hspace{1em}$\beta_{1,1}$ & $-0.000$ & $1.402$ & $0.958$ & $0.033$ & $1.866$ & $0.944$ & $0.039$ & $1.955$ & $0.941$\\
\hspace{1em}$\beta_{0,0}$ & $-0.047$ & $0.704$ & $0.915$ & $0.075$ & $0.853$ & $0.946$ & $0.084$ & $0.887$ & $0.948$\\
\hspace{1em}$\beta_{0,1}$ & $-0.032$ & $1.014$ & $0.920$ & $0.041$ & $1.256$ & $0.934$ & $0.050$ & $1.300$ & $0.931$\\
\hspace{1em}$\beta_{\mathrm{diff},0}$ & $0.130$ & $0.833$ & $0.955$ & $0.092$ & $0.973$ & $0.944$ & $0.104$ & $0.996$ & $0.943$\\
\hspace{1em}$\beta_{\mathrm{diff},1}$ & $0.042$ & $1.253$ & $0.947$ & $0.033$ & $1.460$ & $0.940$ & $0.038$ & $1.503$ & $0.935$\\
\addlinespace[0.3em]
\multicolumn{10}{l}{\textbf{Scenario B: transformed covariates}}\\
\hspace{1em}$\beta_{1,0}$ & $-0.160$ & $1.143$ & $0.821$ & $0.088$ & $1.432$ & $0.958$ & $0.131$ & $1.503$ & $0.949$\\
\hspace{1em}$\beta_{1,1}$ & $-0.062$ & $1.560$ & $0.816$ & $0.040$ & $2.006$ & $0.952$ & $0.058$ & $2.112$ & $0.939$\\
\hspace{1em}$\beta_{0,0}$ & $0.068$ & $0.790$ & $0.960$ & $0.093$ & $0.945$ & $0.954$ & $0.128$ & $0.941$ & $0.950$\\
\hspace{1em}$\beta_{0,1}$ & $0.014$ & $1.162$ & $0.962$ & $0.028$ & $1.520$ & $0.957$ & $0.053$ & $1.425$ & $0.948$\\
\hspace{1em}$\beta_{\mathrm{diff},0}$ & $-0.663$ & $0.943$ & $0.703$ & $0.064$ & $1.176$ & $0.934$ & $0.100$ & $1.144$ & $0.940$\\
\hspace{1em}$\beta_{\mathrm{diff},1}$ & $-0.158$ & $1.405$ & $0.785$ & $0.053$ & $1.866$ & $0.947$ & $0.063$ & $1.730$ & $0.939$\\
\bottomrule
\end{tabular}%
}

\vspace{0.4em}

\parbox{\tablewidth}{\footnotesize
\textit{Note:} The true values of $\beta_{1,0}$, $\beta_{1,1}$, $\beta_{0,0}$, $\beta_{0,1}$, $\beta_{\mathrm{diff},0}$, and $\beta_{\mathrm{diff},1}$ are $-$5.77, 20.85, $-$3.93, 11.27, $-$1.83, and 9.57, respectively. The relative bias (RelBias) is calculated as the median of $(\widehat{\Delta}-\Delta)/\Delta$ over all simulation replications. The MCSD and Coverage denote Monte Carlo standard deviation and 95\% confidence interval coverage, respectively. The confidence intervals of moment-type estimator are constructed based on the nonparametric bootstrap approach, and the Wald-type confidence intervals of the EIF-based estimators are calculated based on proposed variance estimator with normal quantiles.
}
\end{table}

We conduct additional simulation studies with smaller and larger sample sizes, $n=500$ and $n=2{,}000$, respectively. The results are reported in Web Tables~S1--S2 of the Supplementary Material. Overall, the performance patterns are broadly consistent with those observed for $n=1{,}000$. As expected, the Monte Carlo standard deviations decrease as the sample size increases. However, the moment-type estimator continues to exhibit poor coverage under Scenario B when nuisance models are misspecified, and this undercoverage becomes more pronounced in larger samples. In Web Appendix~1.3, we also report simulation results for the group-based ETEs. The findings are qualitatively similar to those for the projection parameters, with the EIF-based estimators showing greater robustness to first-stage nuisance estimation error than the moment-type estimator.

\section{Relation to principal stratification analysis}\label{sec:connections}

In this section, we first discuss the conceptual distinction between empirical stratification and principal stratification. We then present a mathematical relationship between their target causal effects, demonstrating an inherent connection between these two stratification ideas. Under the principal stratification framework, the principal causal effect is defined as
$$
\mathrm{PCE}_{\bm g}=\E[Y(1)-Y(0)\mid \bm G=\bm g],
$$
where $\bm g=[m_1,m_0]^\top$ denotes a target principal stratum in the support of $\bm G=[M(1),M(0)]^\top$. 
The key distinction between principal and empirical stratification is the basis on which individuals are stratified. 
Principal stratification uses the true joint potential values of the PV, $\bm G$, whereas empirical stratification uses baseline predictions of these potential PV responses. 
Consequently, the two frameworks address different scientific questions. 
Principal stratification is most natural when the goal is to understand treatment effects within latent counterfactual response types defined by $\bm G$, although identification of PCEs often requires additional assumptions. 
In contrast, empirical stratification is designed to characterize treatment effect heterogeneity across identifiable subgroups constructed from baseline covariates. By stratifying individuals according to baseline-predicted PV responses, empirical stratification targets subgroups that can, in principle, be described before treatment assignment and may therefore be more directly useful for policy and practice. Table~\ref{tab:ps_vs_es} summarizes the differences in interpretation, identification, and scientific meaning between the two frameworks.

\begin{table}[t]
\centering
\caption{Comparison between the empirical stratification framework and principal stratification framework.}
\label{tab:ps_vs_es}
\footnotesize
\renewcommand{\arraystretch}{1.18}
\setlength{\tabcolsep}{5pt}
\begin{tabular}{
@{}>{\raggedright\arraybackslash}p{0.20\linewidth}
   >{\raggedright\arraybackslash}p{0.38\linewidth}
   >{\raggedright\arraybackslash}p{0.38\linewidth}@{}}
\toprule
&
\textbf{Empirical stratification}
&
\textbf{Principal stratification}
\\
\midrule

{\bf Target subgroup}
&
Empirical strata defined by baseline predictions of potential PV responses,
$\E[\bm G\mid\bcx]$.
&
Principal strata defined by the joint potential values of the PV, $\bm G$.
\\
\addlinespace[1em]

{\bf Observability}
&
Empirical strata can be determined once baseline covariates are collected.
&
Principal strata are generally latent, because $M(1)$ and $M(0)$ cannot be jointly observed.
\\
\addlinespace[1em]

{\bf Target estimand}
&
ETE: the treatment effect within a given empirical stratum.
&
Principal causal effect: the treatment effect within a given principal stratum.
\\
\addlinespace[1em]
{\bf Identification requirements}
&
The ETE is identifiable under standard treatment ignorability and positivity assumptions.
&
The PCE typically requires additional assumptions   such as monotonicity, exclusion restrictions, principal ignorability, or parametric  models.
\\

\addlinespace[1em]

{\bf Research question in WHO-LARES study}
&
How does the effect of damp housing on depression vary among individuals who are more or less likely to develop dampness related disease under alternative housing conditions?
&
How does the effect of damp housing on depression vary among individuals who would or would not develop dampness related disease under alternative housing conditions?
\\
\bottomrule
\end{tabular}
\end{table}

Besides the conceptual difference, the principal causal effect and ETE are mathematically related when we set the empirical score as the principal score $e_{\bm g}(\bcx):=\mathbb{P}(\bm G=\bm g|\bcx)$. Under this specification, the ETE becomes $\mathrm{ETE}_{\bm g}(t)=\E[Y(1)-Y(0)\mid e_{\bm g}(\bcx)=t]$, which can be interpreted as the treatment effect among subgroup individuals with a $t\times 100\%$ probability of membership  within principal stratum $\bm g$. Notably, the principal score cannot be identified based on Assumptions \ref{assum:igno}--\ref{assum:posi} only. In the binary PV setting with $M\in\{0,1\}$, if the monotonicity assumption ($M_i(1) \geq M_i(0)$ for all individuals) is plausible, then $e_g(\bcx)$ is identified by $e_{\bm g}(\bcx)= h_0(\bcx)$, $h_1(\bcx)-h_0(\bcx)$, and $1-h_1(\bcx)$ for $\bm g\in\{[1,1]^\top,[1,0]^\top,[0,0]^\top\}$, respectively, while the stratum $\bm g = [0,1]^\top$ does not exist \citep{angrist1996identification}. 
Our comparison will focus on the scenario where the principal score is identifiable.

$\mathrm{ETE}_{\bm g}(t)$ and $\mathrm{PCE}_{\bm g}$ are not directly comparable because they target different subpopulations. In particular, the former targets an empirical stratum, a baseline-dependent and more localized stratum, whereas the latter targets the overall principal stratum. Therefore, to unify the subpopulation, we compare the PCE to an average ETE over individuals in the principal stratum. Specifically, given an individual with covariates $\bm X$ from the target principal stratum, let $\mathrm{ETE}_{\bm g}\{e_{\bm g}(\bcx)\}$ be the individualized ETE when the principal score is set to the value under observed individual covariates. 
Then, we define the average ETE among individuals in principal stratum $\bm g$ as
$$
\mathrm{AvgETE}_{\bm g} = \E\left[\mathrm{ETE}_{\bm g}\{e_{\bm g}(\bm X)\}\mid \bm G=\bm g\right].
$$
Remark~{S1}  of Supplementary Material  shows that $\mathrm{AvgETE}_{\bm g} $ can be   identified as
$
\mathrm{AvgETE}_{\bm g} =  {\E[e_{\bm g}(\bcx)\{\mu_1(\bcx)-\mu_0(\bcx)\}]}/{\E[e_{\bm g}(\bcx)]}$ under Assumptions~\ref{assum:igno}--\ref{assum:posi} with identified principal scores, whereas $\mathrm{PCE}_{\bm g}$ is still not identifiable even if under these conditions.
Theorem \ref{thm:linketepce_v2} gives an exact decomposition of the discrepancy between $\mathrm{AvgETE}_{\bm g}$ and $\mathrm{PCE}_{\bm g}$, thereby clarifying when these two population summaries coincide or differ.

\begin{theorem}\label{thm:linketepce_v2}
Assuming $\E[e_{\bm g}(\bcx)] >0$, the difference between $\mathrm{PCE}_{\bm g}$ and $\mathrm{AvgETE}_{\bm g}$ is
\begin{align}\label{thm:linketepce:eq}
\mathrm{PCE}_{\bm g}  - \mathrm{AvgETE}_{\bm g} = \frac{\E[ \delta_{\bm g}^{(1)}(\bm X) \times \delta_{\bm g}^{(2)}(\bm X)]}{\E[e_{\bm g}(\bcx)]},
\end{align}
where $\delta_{\bm g}^{(1)}(\bm X)=e_{\bm g}(\bcx)\{1- e_{\bm g}(\bm X)\}$ and $\delta_{\bm g}^{(2)}(\bm X) = \E[Y(1)-Y(0)\mid \bm X, \bm G =  \bm g] - \E[Y(1)-Y(0)\mid \bm X, \bm G \neq  \bm g]$.
\end{theorem}

The difference between \(\mathrm{PCE}_{\bm g}\) and \(\mathrm{AvgETE}_{\bm g}\) in \eqref{thm:linketepce:eq} is small when, for most individuals, either \(\delta_{\bm g}^{(1)}(\bcx)\) or \(\delta_{\bm g}^{(2)}(\bcx)\) is small.  The first factor, $\delta_{\bm g}^{(1)}(\bcx)=e_{\bm g}(\bcx)\{1-e_{\bm g}(\bcx)\}$, measures the uncertainty
in principal-stratum membership given baseline covariates. 
 When $e_{\bm g}(\bcx)$ is close to either zero or one for most individuals, principal-stratum
membership is nearly determined by $\bcx$, then $\delta_{\bm g}^{(1)}(\bcx)$ becomes small and $\mathrm{AvgETE}_{\bm g} \approx \mathrm{PCE}_{\bm g}$. In the extreme case where $\bm G$ is discrete and fully observable, such that $\bm G$ is included in observed covariates $\bcx$, we have $e_{\bm g}(\bcx)=1$ and $0$ if $\bm G=\bm g$ and $\bm G \neq \bm g$,  respectively. Then $\delta_{\bm g}^{(1)}(\bcx)\equiv 0$ for all $\bm X$, and therefore $\mathrm{PCE}_{\bm g}  = \mathrm{AvgETE}_{\bm g}$. The second factor, $\delta_{\bm g}^{(2)}(\bcx)$, measures the degree to which the principal ignorability assumption is violated. 
In particular, principal ignorability is a widely used assumption for identifying
the PCE
\citep{jo2009use,jo2011use,stuart2015assessing,jiang2021identification}, which
requires \(\{Y(1),Y(0)\} \indep G \mid \bcx\) and is plausible if there are no unmeasured confounders for the $\bm G$--$Y$ relationship\footnote{Weaker versions of
principal ignorability have been further considered in the literature
\citep{ding2017principal,forastiere2018principal}, but  we use its original form
 here  because it enjoys better interpretation.}. Correspondingly, if principal ignorability holds, then $\delta_{\bm g}^{(2)}(\bcx)\equiv 0$ and $\mathrm{AvgETE}_{\bm g}=\mathrm{PCE}_{\bm g}$.

\begin{remark}[Connection to compliance-score adjusted causal effects]
\label{rem:compliance_score}
In randomized trials with noncompliance, let $Z\in\{0,1\}$ denote treatment assignment and let $M\in\{0,1\}$ denote actual treatment receipt. Prior work studies conditional intention-to-treat effects across levels of a compliance score, $\mathbb{P}\{\bm G=[1,0]^\top\mid \bcx\}$, i.e., the probability belonging to the compliers stratum \citep{follmann2000effect,joffe2003compliance,hu2022assessing}. Under the monotonicity assumption $M(1)\ge M(0)$, the empirical score is identified as $\mathbb{P}\{\bm G=[1,0]^\top\mid \bcx\}=h_1(\bcx)-h_0(\bcx)$. Under this specification, the corresponding compliance-score adjusted causal effect can be viewed as an ETE obtained by choosing the empirical score $\tau(\bcx)$ to be this compliance score. 
Thus, empirical stratification generalizes compliance-score analysis from treatment noncompliance to broader classes of PVs.
\end{remark}

\section{Two applications}\label{sec:application}

\subsection{The WHO-LARES study with an intercurrent event}\label{sec:application_wholares}

We apply the proposed methodology to the WHO-LARES study ($n=5{,}882$) to study how the effect of living in damp housing on depression varies with individuals' vulnerability to dampness-related disease. We define the empirical score as $\tau(\bcx)=s_1(\bcx)=\Prob\{M(1)=1\mid \bcx\}$, so that individuals with larger values of $\tau(\bcx)$ are those predicted to be more vulnerable to dampness-related disease if exposed to damp housing. We focus on the projected ETE curve under the linear working model so that $\mathrm{projETE}(t)=\beta_{\mathrm{diff},0}+\beta_{\mathrm{diff},1}t$. The nuisance parameters $\{\pi_z(\bcx),h_z(\bcx),\mu_z(\bcx)\}$ are estimated using either parametric models or machine learning methods, leading to two EIF-based estimators,  $\widehat{\mathrm{projETE}}^{\mathrm{eff-par}}(t)$ and $\widehat{\mathrm{projETE}}^{\mathrm{eff-ml}}(t)$. The nuisance-model specifications follow those used in Section~\ref{sec:simulation}. To make the treatment ignorability assumption more plausible, we collect the following baseline covariates: sex, age, marital status, education, employment, smoking, home ownership, home size, crowding (number of residents per room), heating, and natural light.

Before presenting the empirical treatment effect analysis, we first examine how the estimated empirical score relates to baseline covariates. Web Table S3 in the Supplementary Material summarizes baseline characteristics for individuals below and above the median of the estimated empirical score, where $\widehat\tau(\bcx)$ is obtained using the machine learning nuisance estimator. This descriptive comparison  characterizes which covariates influence individuals' vulnerability to dampness-related disease. As shown in Web Table S3, a particularly notable pattern is the sharp difference in gender composition across the two groups: the lower-score group is composed mostly of men, whereas the higher-score group is composed predominantly of women. In addition, individuals in the higher-score group tend to be older, less likely to be married or employed, and more likely to smoke. They are also less likely to own their homes and more likely to live in crowded dwellings. Overall, the higher-score group is characterized by less favorable baseline profiles, suggesting greater predicted vulnerability to dampness-related disease under damp housing exposure.

\begin{figure}[t]
    \centering
    \includegraphics[width=0.9\textwidth]{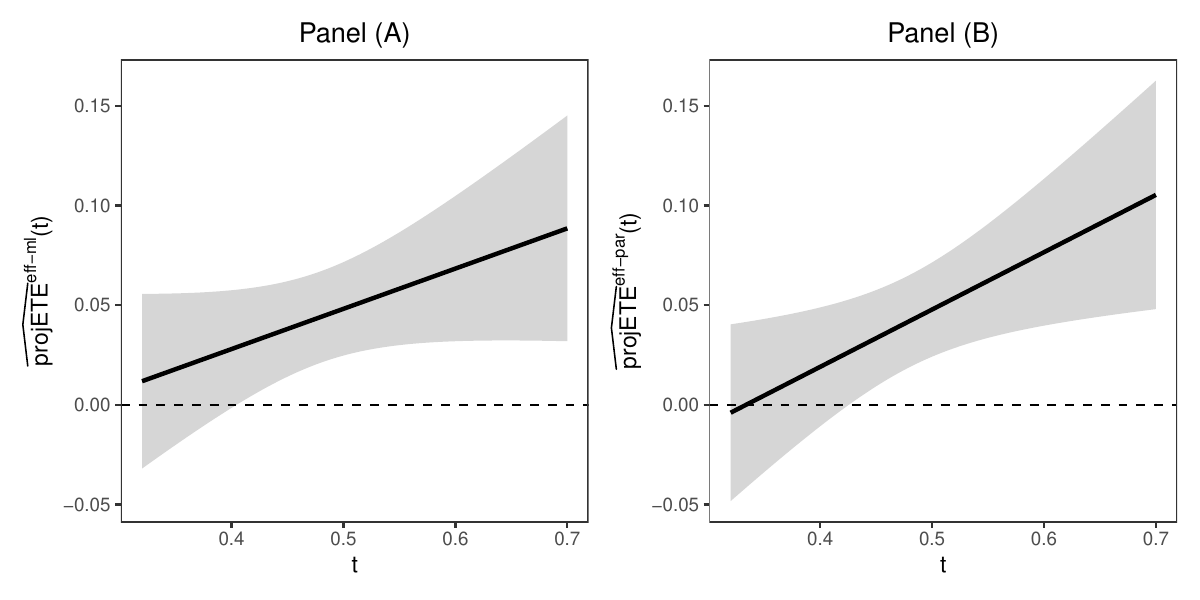}
    \caption{Projected ETE curves in the WHO-LARES study. Panel (A) shows the EIF-based estimator using machine learning nuisance models, and Panel (B) shows the EIF-based estimator using parametric nuisance models. The solid line represents the estimated projected ETE curve, and the shaded region represents the pointwise 95\% confidence band.}
    \label{fig:projETE_application}
\end{figure}

Figure~\ref{fig:projETE_application} displays the projected ETE curve over the empirical score range from 0.32 to 0.70, which corresponds to the 2.5th to 97.5th percentiles of the  empirical score distribution under the machine learning estimator. Specifically, using the machine learning estimator, the estimated coefficients for the linear working model are $\widehat{\beta}_{\text{diff},0}^{\mathrm{eff-ml}}=-0.053$ (SE $=0.059$) and $\widehat{\beta}_{\text{diff},1}^{\mathrm{eff-ml}}=0.202$ (SE $=0.120$). The positive estimate of $\beta_{\text{diff},1}$ indicates that the adverse effect of living in damp conditions on depression becomes stronger as the empirical score increases, i.e., among individuals with greater vulnerability to dampness-related disease. Quantitatively, a 0.1-unit increase in the empirical score corresponds to an increase of about $2.02\%$ in the projected effect of living in damp conditions on depression. The corresponding results based on parametric nuisance models are qualitatively  similar with $\widehat{\beta}_{\text{diff},0}^{\mathrm{eff-par}}=-0.096$ (SE $=0.059$) and $\widehat{\beta}_{\text{diff},1}^{\mathrm{eff-par}}=0.287$ (SE $=0.121$).

{\color{black}From a policy perspective, the positive trend in $\mathrm{projETE}(t)$ suggests that
individuals with higher empirical scores are both more vulnerable to
dampness-related disease and depression due
to damp housing. Thus, policymakers may prioritize interventions to improve damp living conditions for individuals with higher
empirical scores, because they are expected to experience greater mental and physical health gains from such interventions. Such policy insight is less directly available
from conventional HTE analyses based only on baseline covariates, which cannot
characterize treatment effect heterogeneity for the PV and the outcome simultaneously. It is also actionable since empirical stratum membership is observed solely
based on baseline covariates.}

As a secondary analysis, we estimate group-based ETEs (Remark \ref{remark:grpETE}) for individuals below and above $\tau(\bcx)=0.51$, which is the median of the estimated empirical score by the machine learning estimator. This allows us to calculate two grpETEs, with $\text{grpETE}(1)$ and $\text{grpETE}(2)$ corresponding to the treatment effect among the lower and higher empirical score subgroups, respectively. Detailed inference procedures for calculating the group-based ETEs are provided in Web Appendix 1.2 of the Supplementary Material. Table~\ref{tab:app_grpete_est} reports the group-based ETE estimates from two EIF-based estimators, one using machine learning nuisance models and the other using parametric nuisance models. The two estimators give very similar results. Under the machine learning estimator, the estimated group-based ETE is 0.037 for the lower-score group and 0.058 for the higher-score group, suggesting that living in damp housing has a more harmful effect on depression among individuals with greater predicted vulnerability to dampness-related disease. These group-based estimates are consistent with the positive slope shown in the projected ETE curve.

\begin{table}[t]
\centering
\caption{\label{tab:app_grpete_est}Estimation results for group-based ETEs, WHO-LARES study.}
\vspace{-0.4cm}
\centering
\resizebox{\ifdim\width>\linewidth\linewidth\else\width\fi}{!}{
\fontsize{10}{12}\selectfont
\begin{tabular}[t]{lcccccc}
\toprule
\multicolumn{1}{c}{ } & \multicolumn{3}{c}{Machine learning estimator} & \multicolumn{3}{c}{Parametric estimator} \\
\cmidrule(l{3pt}r{3pt}){2-4} \cmidrule(l{3pt}r{3pt}){5-7}
Estimand & $\mathcal T_k$ & Estimate & 95\% CI & $\mathcal T_k$ & Estimate & 95\% CI\\
\midrule
grpETE(1) & $[0.271, 0.510]$ & $0.037$ & $[0.006, 0.068]$ & $[0.276, 0.510]$ & $0.034$ & $[0.004, 0.064]$\\
grpETE(2) & $(0.510, 0.800]$ & $0.058$ & $[0.023, 0.093]$ & $(0.510, 0.790]$ & $0.060$ & $[0.024, 0.095]$\\
\bottomrule
\end{tabular}}
\end{table}

\subsection{The National Job Corps study with  noncompliance}

The National Job Corps Study is a randomized trial of the Job Corps training program for disadvantaged youths in the United States \citep{schochet2001national}, where participants were assigned either to have immediate access to Job Corps  ($Z=1$) or to a control condition in which access to Job Corps was restricted for
three years ($Z=0$). We study the effect of assignment to Job Corps on participants' weekly earnings during the fourth year after randomization ($Y$). Treatment noncompliance is substantial in the study \citep{frumento2012evaluating,chen2015bounds}. We define $M=1$ if an individual participated in Job Corps within one year after randomization and $M=0$ otherwise. Among the 9,240 participants, 15\% of individuals assigned to the treatment group did not participate in Job Corps, while about 51\% of individuals assigned to the control group participated in the program. 

We define the empirical score as $\tau(\bcx)=s_1(\bcx)-s_0(\bcx)$, so that
$\tau(\bcx)$ summarizes the extent to which an individual's treatment receipt is
affected by treatment assignment. In the National Job Corps Study, the
monotonicity assumption is plausible, because the Job Corps program is generally
considered beneficial and defiers are unlikely to exist \citep{chen2015bounds}.
As discussed in 
Remark~\ref{rem:compliance_score}, $\tau(\bcx)$ is the compliance score under monotonicity. Therefore, we shall assume the monotonicity assumption and interpret $\mathrm{ETE}(t)$ as the compliance-score adjusted causal
effect, measuring the treatment effect among individuals with a $t\times 100\%$
predicted probability of belonging to the complier stratum.

Although the National Job Corps Study is a randomized trial, we consider a rich set of baseline covariates when estimating the nuisance parameters (see Web Table S4 in the Supplementary Material for the list of covariates). The covariates are used to improve the prediction of treatment receipt behavior in order to construct a more informative compliance score. To describe how compliance scores relate to baseline covariates, Web Table~S4 in the Supplementary Material summarizes baseline characteristics across two subgroups with individuals below and above the median of the compliance score distribution. Individuals in the higher-score group appear to have better socioeconomic status, who are older, more educated, and more likely to have a high school degree or GED degree. They also have higher baseline earnings and are more likely to report smoking or drinking experience. 

\begin{figure}[t]
    \centering
\includegraphics[width=0.9\textwidth]{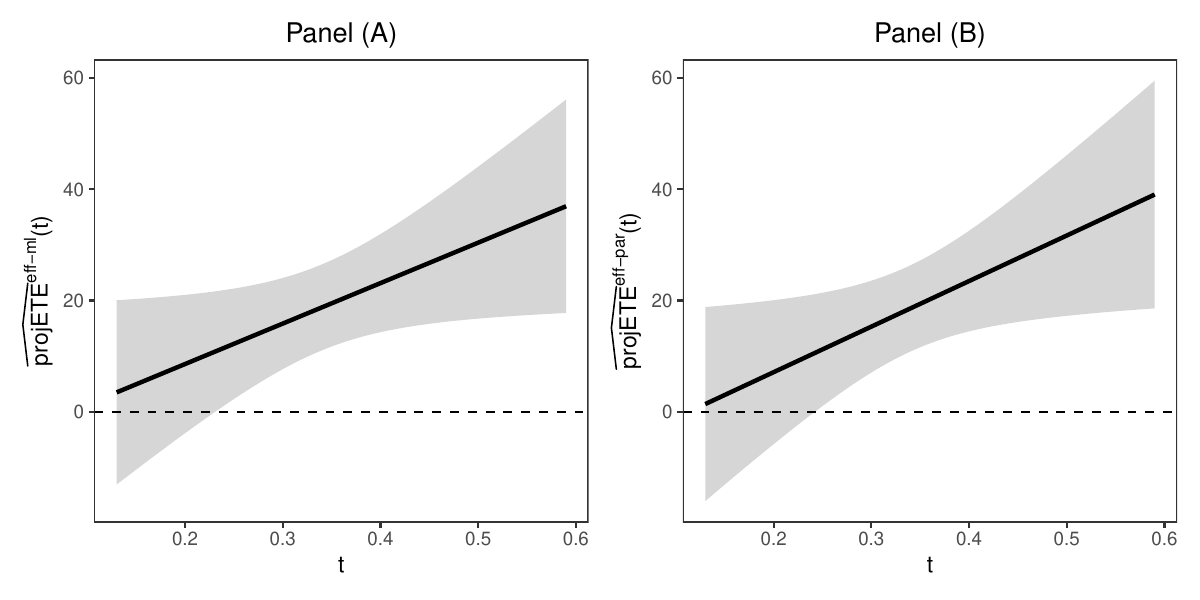}
    \caption{Projected ETE curves in the National Job Corps study. Panel (A) shows the EIF-based estimator using machine learning nuisance models, and Panel (B) shows the EIF-based estimator using parametric nuisance models. The solid line represents the estimated projected ETE curve, and the shaded region represents the pointwise 95\% confidence band.}
    \label{fig:projETE_noncompliance}
\end{figure}

As in Section~\ref{sec:application_wholares}, we estimate the projected ETE curve under a linear working model,
so that $\mathrm{projETE}(t)=\beta_{\mathrm{diff},0}+\beta_{\mathrm{diff},1}t$.
We consider two EIF-based estimators, where one uses parametric nuisance models
($\widehat{\mathrm{projETE}}^{\mathrm{eff}\text{-}\mathrm{par}}(t)$) and the other uses
machine learning nuisance models
($\widehat{\mathrm{projETE}}^{\mathrm{eff}\text{-}\mathrm{ml}}(t)$). The results are
shown in Figure~\ref{fig:projETE_noncompliance}. Both estimators show an upward-sloping
pattern. The estimated slope is
$\widehat\beta^{\mathrm{eff}\text{-}\mathrm{par}}_{\mathrm{diff},1}=81.88$
(SE $=38.35$) using parametric nuisance models and
$\widehat\beta^{\mathrm{eff}\text{-}\mathrm{ml}}_{\mathrm{diff},1}=72.73$
(SE $=35.73$) using machine learning nuisance models. 
Quantitatively, based on the machine learning estimator, a 0.1-unit increase in the compliance score is associated with an increase of about 7.27 dollars in weekly earnings. Therefore, the economic benefit of assignment to Job Corps is concentrated among individuals whose participation behavior is more responsive to treatment assignment.

The estimated ETE curves in Figure~\ref{fig:projETE_noncompliance} are not  negative over the range of $\tau(\bcx)$. This suggests that the access to Job Corps has a nonnegative causal effect on future earnings among disadvantaged youth across the entire range of compliance scores.   Individuals with higher compliance scores are both more likely to participate in
Job Corps and have larger estimated gains
in weekly earnings. Policymakers may therefore use the compliance score to
inform program targeting and implementation: individuals with higher compliance scores may be prioritized when resources are limited. Individuals with lower compliance scores may require additional support to convert program access into actual participation; they may also benefit from a more tailored training program for a better improvement of their future earnings.

\section{Discussion}\label{sec:discussion}



In this paper, we mainly consider settings in which the PV is distinct from the primary outcome. 
In some applications, predictive HTE analyses do not involve a separate PV and instead stratify individuals by their predicted outcome under control to examine whether treatment effects vary across outcome-based risk groups \citep{kent2020predictive}. 
This setting can be viewed as a special case of our framework by taking the PV to be the primary outcome itself and specifying the empirical score as $s_0(\bcx)$. 
The proposed estimation and inference procedures can therefore also be applied to the conventional outcome-based predictive HTE analysis.

Several limitations and extensions remain. First, as in most causal inference methods, identification of the ETE relies on treatment ignorability assumption, which may not hold in observational studies. In applications such as the WHO-LARES study, where treatment is not randomized, then the analysis may be biased due to unmeasured confounding. Sensitivity analyses for violations of treatment ignorability are a useful extension. Second, the projected ETE curve summarizes treatment effect heterogeneity through a prespecified working model. While this improves stability and interpretability, it may obscure more complex nonlinear patterns in the underlying ETE curve. If $\tau(\bcx)$ is continuously distributed, one possible extension is to develop kernel smoothing estimators that locally average the ETE curve over neighborhoods of the empirical score \citep{zhang2025semiparametric}. This would provide a more flexible, nonparametric description of how treatment effects vary with predicted PV responses. Third, many applications involve multiple PVs. For instance, in pragmatic trials, investigators may observe both treatment noncompliance and intercurrent events (e.g., the ADAPTABLE trial \citep{jones2021comparative}). Extending empirical stratification to this setting would require defining empirical strata based on the joint predicted response profile of multiple PVs, such as $\E[\bm M(1)\mid \bcx]$ and $\E[\bm M(0)\mid \bcx]$ for a vector of PVs $\bm M$. Such an extension would allow investigators to characterize treatment effect heterogeneity driven jointly by multiple PVs.

\section*{Supplementary material}
Supplementary Material includes technical supporting information, the semiparametric estimation framework for the group-based ETE, and additional simulation studies.

\section*{Acknowledgements}

We thank the World Health Organization’s European Centre for Environment and Health, Bonn office, for providing the WHO-LARES data. We thank Peng Ding and Fan Li for their helpful feedback, which helped improve the manuscript.

\spacingset{1}

\bibliographystyle{jasa3}

\bibliography{Bibliography-MM-MC}

\end{document}